\documentclass[a4paper,10pt]{amsart}
\usepackage{graphicx}
\begin{document}

\title[]{FCS diffusion laws on two-phases \\lipid membranes : experimental and Monte-carlo simulation determination of domain size.}%
\author{\tiny{C.Favard}}
\author{J. Ehrig}
\author{J. Wenger}
\author{P.-F. Lenne}
\author{H. Rigneault}
\address{ Mosaic Project, Institut FRESNEL, CNRS UMR6133, Campus de Saint J\'{e}r\^{o}me, av. Escadrille Normandie Niemen, Marseille, 13013, France.}%
\email{cyril.favard@fresnel.fr}%

\thanks{}%
%\subjclass{}%

\date{May 3rd, 2010}%
%\dedicatory{}%
%\commby{}%
% ----------------------------------------------------------------
\begin{abstract}
For more than ten years now, many efforts have been done to identify
and characterize nature of obstructed diffusion in model and
cellular lipid membranes. Amongst all the techniques developed for
this purpose, FCS, by means of determination of FCS diffusion laws,
has been shown to be a very efficient approach. In this paper, FCS
diffusion laws are used to probe the behavior of a pure lipid and a
lipid mixture at temperatures below and above phase transitions,
both numerically, using a full thermodynamic model, and
experimentally. In both cases FCS diffusion laws exhibit deviation
from free diffusion and reveal the existence of domains. The
variation of these domains mean size with temperature is in perfect
correlation with the enthalpy fluctuation.
\end{abstract}
\maketitle
% ----------------------------------------------------------------

\section{Introduction}

Since the mosaic fluid concept of Singer and Nicolson \cite{Sin72}
where lipids were considered as a "sea" in which protein were
embedded, the description of biological lipid membranes has evolved
to a spatio-temporal heterogeneous mixture of its own components. It
is now mainly admitted that biological membranes are organized in
domains of different compositions and different size including nano,
meso and microscopic scale organisation. Membrane heterogeneity may
be of various types. Several lipid lamellar phases have been
identified in model systems. Basically, lipid bilayer can exist in
solid phases ($s$ also named gel phases $g$), liquid disordered
phases ($l_{d}$ also named fluid phases $f$), and liquid ordered
phases ($l_{o}$) which are often enriched in cholesterol. In complex
lipid mixture, such as biological membranes, coexistence of these
different phases may occur, leading to the formation of domains
\cite{Bro98}.This led to the concept of \textit{raft} as functional
domains existing within biological membranes as reviewed by Simon
and Ikonen \cite{Sim97} that highlighted their potentially
ubiquitous role in cell biology. A lot of work has been done on
rafts for 10 years now \cite{Bro00, Muk04, Sim04, Jac07} that led to
the present consensus that rafts are heterogeneous membrane
structures, rich in cholesterol and sphingomyelin and about 10 to
200 nm in diameter, highly dynamic in the lipid membranes of all
eukaryotic cells \cite{Pik06}. Since this size is below the
diffraction limit there are no direct images of their existence.
Moreover, the big discrepancy founded in size mainly depends on the
technique used to reveal them. For example, donor quenching FRET
analysis shows nanoscale domain formation (10 to 40 nm) in lipid
bilayers with a similar composition to that of the outer plasma
membrane, at temperature of 37° C, whereas macroscopic phase
separation was not evident \cite{Sil03}. Similarly, $l_{o}$
nanodomains can be detected by FRET in regions of the phase diagram
in which confocal microscopy indicates only the presence of a single
homogenous phase \cite{Fei01}. Domains on length scales of 30 to 80
nm in a region of Lo and Ld co-existence have also been detected
using atomic-force microscopy (AFM)\cite{Yua02}, deuterium-based
nuclear magnetic resonance (2H-NMR) and differential scanning
calorimetry \cite{Hsu02,Vea04}. Beside their absolute size, since
existence of domain in complex mixtures is dynamic, it seems
relevant to use diffusion as a spatio-temporal probe of the local
environment. This approach has been developed through Fluorescence
Recovery After Photobleaching (FRAP) \cite{Alm92,Vaz89,Vaz91} or
Fluorescence Correlation Spectroscopy (FCS) \cite{Kor02, Sch09,
Fei99, Egg09} and Single Particle Tracking (SPT)\cite{Die02}.

\begin{figure}[!h]
   \begin{center}
      \includegraphics*[width=4in,height=3in]{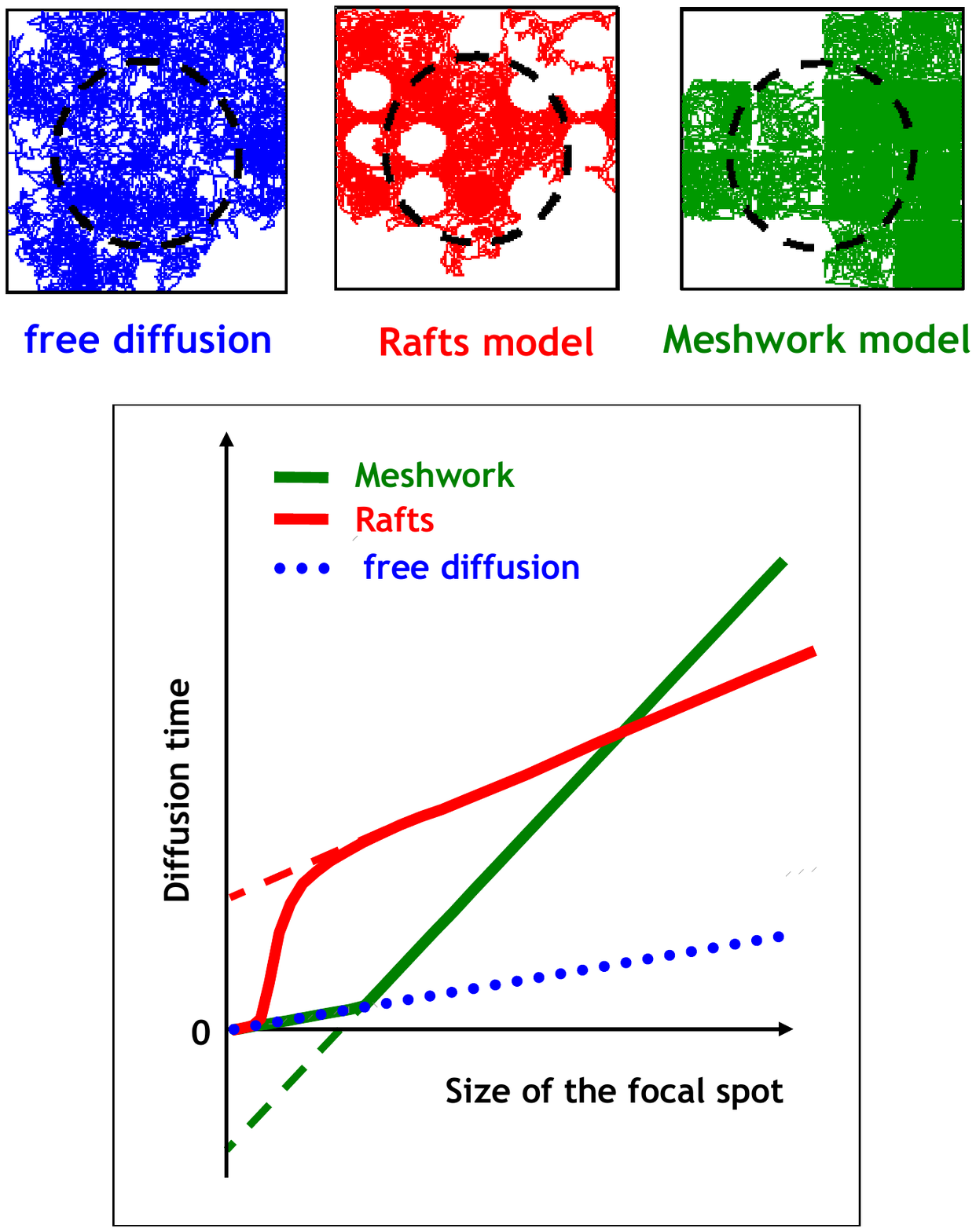}
      \caption{\small{\textbf{FCS diffusion laws :}}
      \scriptsize{FCS diffusion laws are obtained by plotting variation of the FCS
diffusion time as a function of the size of the focal spot (or
waist). Here are illustrated three typical different diffusion laws
obtained by numerical simulation. One for free diffusion (blue line)
and two for obstructed diffusion by raft (red line) or meshwork
(green line). For both obstructed diffusion models, it can be seen
that FCS diffusion laws are not linear and that their asymptotic
behavior can be fitted using affine model (For detailed explanations
see \cite{Waw05}).}}
      \label{fig1}
   \end{center}
\end{figure}

Each of the dynamic microscopic techniques cited above have typical
advantage and disadvantages in respect to their timescale and
statistics. For example, FCS is sensitive on the millisecond to
second timescale, corresponding to diffusion characteristic time in
lipid mixtures of a fluorescently labeled molecule through the focus
with a waist of approximately 200nm. Recently, Wawrezinieck
\textit{et al.} \cite{Waw05} showed that FCS was a powerful tool for
analyzing complex diffusions. They showed that direct fitting of the
autocorrelation function was unable to discriminate between these
complex diffusions but on the opposite, exploration at different
space scale allowed a more detailed view of the environment
structure. They developed a variable waist FCS experiment and, based
on pure numerical simulation they succeeded in reassigning different
type of FCS diffusion laws as a function of the probed environment
such as \textit{rafts} or meshwork (see Fig~\ref{fig1}). This was
furthermore confirmed by measuring FCS laws in living cells
\cite{Len06} and use recently at different space scale down to 50 nm
\cite{Wen07, Egg09}. Even though extensive numerical simulation in
given geometries showed a fairly accurate prediction to what
%in the models were based on pure Monte-Carlo simulations of a
%tracer in fairly defined geometrical environments and applied to
occurs in very complex systems such as cell membranes, a level of
organisation in between is still missing. Therefore, investigations
were conducted in order to further explore the ability of these
quantitative FCS diffusion laws to distinguish phases during the
transition in a well defined lipid mixture such as DMPC-DSPC.
%test the ability for FCS diffusion law to quantitatively
%probe the environment it seemed important to test the ability of

DMPC/DSPC lipid mixtures has been studied extensively for a while.
The thermodynamic parameters of this mixture have been deciphered by
many different techniques including differential scanning
calorimetry, neutron scattering, NMR, ESR, Raman spectroscopy and
Fourier transform infrared spectroscopy \cite{See08} or AFM
\cite{Gio04}. The structural characteristics of this mixture has
also been approached by FRAP \cite{Vaz89,Sch96}. These studies have
shown that DMPC-DSPC form non-ideal two-phases mixtures with small
clusters of the minor phase in a continuum of the major phase.
Determination of thermodynamic parameters also allowed to develop
numerical simulations of the lipid mixture. Monte Carlo methods has
been used to simulate the lateral distribution of each component in
the pure gel or fluid phase of DMPC/DSPC mixture, firstly assuming
one state and two components \cite{Jan84} and more recently
computing a two-phases two-components triangular lattice model
\cite{Sug99}. This model has been thoroughly used to predict size,
shape and number of gel($g$)/fluid($f$) clusters \cite{Sug05} and
has recently been used to simulate FCS experiments on it
\cite{Hac05}.

The accuracy of the thermodynamic microscopic description of
DMPC:DSPC mixtures provides the ideal model system to explore the
relevancy of FCS diffusion laws. It is the scope of this paper to
extend the previous works \cite{Jan84,Sug99,Sug05,Hac05} into this
direction both theoretically and experimentally. Here are reported
FCS diffusion laws on both DMPC:DSPC (8:2 mol:mol) mixtures and pure
DMPC at different temperatures spanning the range of the phase
transitions. These FCS laws shows deviations from a pure Brownian
motion as expected. In order to understand the origin of this
deviation and quantitatively characterize it, numerical simulations
have been performed on these lipids. The results found here show
that FCS diffusion laws are able to distinguish the existence of
domains and that the mean size of the domains can be determined from
it. A simple model is given to explain the shape of the FCS
diffusion laws found here as a function of diffusion in a solid or a
liquid environment.
%DMPC-DSPC mixtures are numerically simulated at different temperature and FCS diffusion laws are determined from these simulations. It is shown that these FCS diffusion laws are able to distinguish the existence of domains within the simulation and that the mean size of the domains can be determined from it. Then, FCS diffusion laws are experimentally made on multilamellar DMPC-DSPC lipid mixture within the same temperature range. These experimental FCS laws show correct agreement with the numerical results, allowing the determination of the mean size of the domains in the sample. Finally, an simple model is given to explain the shape of the FCS diffusion laws found here as a function of diffusion in solid or liquid environment and fluctuation of the lipid mixture.

\section{Material and Methods}
\subsection{Material}
The lipids 1,2-distearoyl-\textit{sn}-glycero-3phosphocoline (DSPC)
and 1,2-dipalmitoyl-\textit{sn}-glycero-3-phospho-choline (DMPC)
were purchased from Avanti Polar Lipids (Alabaster, AL.). They were
used without further purification and kept at $-20$\r{°}C in
chloroform:methanol(9:1, mol:mol) at 100mM concentration. For FCS
measurements lipids mixtures were labeled with
2-(4,4-difluoro-5,7-dimethyl-4-bora-3a,4a-diaza-s-indacene-3-pentanoyl)-
1-hexadecanoyl-\textit{sn}-glycero-3-phospho-\\choline($\beta$-BODIPY
FL C5-HPC)from Invitrogen (Carlsbad, CA, USA)).

\subsection{Monte Carlo simulation of the two phases two components lipid mixture}
Our Monte Carlo simulations were directly adapted from the work of
Sugar \textit{et al.} \cite{Sug99} also described in \cite{Hac05}.
Briefly, the thermal fluctuations of the DMPC : DSPC lipid mixture
was simulated using a two-state Ising type monolayer triangular
lattice. Each lattice point is occupied by one acyl chain of either
DMPC or DSPC.
During the simulation,trial configuration are generated by means of six different elementary steps :\\
- One that can be described as a \textbf{phase transition step}, consists in changing the state of a randomly selected acyl chain from gel to fluid or inversely.\\
- Five that can be described as \textbf{diffusion steps}, consist in
exchanging two neighboring molecules. In the Monte Carlo algorithm,
3 different processes can occur at 3 different time scales. One is
diffusing in a liquid environment, the other is diffusing in gel
environment and the last is changing its state.
%Since these processes are different each of them has to be given different frequencies in the Monte-Carlo algorithm.
For lipids within a mixed environment (both gel and fluid) the
probability of entering a diffusion step depends on the fraction of
gel lipids chains ($f_{c,g}$) surrounding the two lipids that are to
enter a diffusion step. As extensively described in \cite{Hac05}, a
rate function $r(f_{c,g})$ has been introduced in the simulation :
\begin{equation} \label{rate}
   r=r_{0}exp(-f_{c,g}\frac{\Delta E}{kT})
   \end{equation}
where $r_{0}$ is the value of $r$ in a fully liquid environment
($f_{c,g}=0$) and $\Delta E$ is the energy barrier needed for a
diffusion step in an all-gel environment ($f_{c,g}=1$). As in the
work of Hac \textit{et al.} \cite{Hac05}, the phase transition step
probability ($r_{state}$) was set to be equal to $r_{0}$ and $\Delta
E/kT$ was set to 4.25 according to the following ratio :
$\frac{D_{f}}{D_{g}}=70$

The Monte-Carlo simulations were implemented in C++ (Microsoft
Visual C++ Version 6.0). They were run on a multiprocessor computer
(4 AMD opterons dual-core 865 at 1.8 Ghz) using the Intel C++
compiler (icc v10.0). For each type of simulations periodic boundary
conditions were used. For the lipid mixture, simulations were made
on 8:2 mol:mol DMPC-DSPC lipid mixtures at different
temperatures(294, 298, 299, 300, 301, 302, 304, 306, 308, 310, 315
and 325 K). A 60x60 lipids lattice (60x120 lipid chains) (1 lipid=1
lattice unit square) was simulated. For DMPC alone, a 40x40 lipid
lattice (40x80 lipid chains) was used. $10^{8}$ Monte Carlo steps
were made for each simulation except at 294 K (1.2. $10^{8}$).

\subsubsection{Thermodynamic model}
The thermodynamic model used here is extensively described in
\cite{Sug99} and \cite{Hac05} Briefly, each lipid chains can exist
in two states, gel($g$) of fluid($f$), these states being different
in internal energy as well as in entropy. Therefore the total energy
of one layer of the lipids in a given configuration C is given by :
\begin{eqnarray} \label{eqn:lipidEne}
 E(C)=n^{g}_{A}E^{g}_{A}+n^{f}_{A}E^{f}_{A}+n^{g}_{B}E^{g}_{B}+n^{l}_{B}E^{l}_{B} \nonumber \\
 +n^{gg}_{AA}E^{gg}_{AA}+n^{gf}_{AA}E^{gf}_{AA}
 +n^{ff}_{AA}E^{ff}_{AA}+n^{gg}_{AB}E^{gg}_{AB}\nonumber \\
 +n^{gf}_{AB}E^{gf}_{AB}+n^{fg}_{AB}E^{fg}_{AB}+n^{ff}_{AB}E^{ff}_{AB}+
 n^{gg}_{BB}E^{gg}_{BB}+n^{gf}_{BB}E^{gf}_{BB}+n^{ff}_{BB}E^{ff}_{BB}
\end{eqnarray}
where $n^{g}_{A,B}$ and $n^{f}_{A,B}$ are the numbers of the gel and
fluid lipid chains and the $E^{g}_{A,B}$ and $E^{f}_{A,B}$ terms are
the respective internal energies of the two states of species A and
B. It is shown in \cite{Sug99} that the Gibbs free energy for a
given state can be written as :
\begin{eqnarray} \label{eqn:lipEnlib}
G = G_{0}+n^{f}_{A}(\Delta H_{A}-T\Delta S_{A})+n^{f}_{B}(\Delta H_{B}-T\Delta S_{B})\nonumber \\
 +n^{gf}_{AA} \omega ^{gf}_{AA}+n^{gf}_{BB} \omega ^{gf}_{BB} +n^{gg}_{AB} \omega ^{gg}_{AB} +n^{ff}_{AB} \omega ^{ff}_{AB} +n^{gf}_{AB} \omega ^{gf}_{AB}+n^{fg}_{AB} \omega ^{fg}_{AB}
\end{eqnarray}

Where $\Delta H_{A}$ and $\Delta H_{B}$ the calorimetric enthalpies,
$\Delta S_{A}= \frac{\Delta H_{A}}{T_{m,A}}$ and $\Delta S_{B}=
\frac{\Delta H_{B}}{T_{m,B}}$ are the respective melting entropies
with $T_{m,A}$ and $T_{m,B}$ being the melting temperatures of the
two pure components and $G_{0}$ the energy of an all gel lipid
matrix. All the other parameters ($\omega ^{\alpha \beta}_{ij}$) are
the nearest-neighbor interaction parameters of a lipid chain species
$i$ in state $\alpha$ with a lipid chain species $j$ in state
$\beta$. The whole set of parameter described above has been given
values according to \cite{Hac05} since these values has been
obtained on multilamellar vesicles. These values are summarized in
table~\ref{table1}.

\subsubsection{FCS diffusion laws in the simulation}
In order to build FCS diffusion laws in the MC simulations, 20$\%$
of the total number of DMPC lipids chains were introduced as
markers. During the simulation, these lipids move and generate
intensity according to their position in the laser profile which is
considered as a Gaussian with a given waist $w$(value of the radius
at $1/e^2$ intensity),

\begin{equation} \label{eqn:intensgen}
   I(r)\propto exp(-\frac{2(r-r_{0})^2}{w^2})
\end{equation}

At each time step, the detected intensity in our simulations is
computed assuming a Poisson distribution. The number of detected
photons ($n_{ph}$) for a particle at position (x,y) is given by a
random variable following the Poisson distribution with parameter
$\beta$I(x,y) with $\beta$ describing the collection efficiency of
the setup \cite{Woh01}.

To analyse intensity fluctuations, the normalized time
autocorrelation function (ACF) is defined as :
\begin{equation} \label{eqn:normACF}
   g^{(2)}(\tau) = \frac{< n_{ph}(t)n_{ph}(t+\tau)>}{<n_{ph}(t)>^{2}}
\end{equation}
where $<.>$ represent a time average. In our simulations, the ACF is
calculated from a generated intensity file after the whole Monte
Carlo simulation. The software correlator used to compute the ACFs
follows the architecture proposed by \cite{Sch85} and described in
\cite{Woh01}. It has a logarithmic timescale, each channel having an
individual sampling time and delay time.

FCS diffusion laws are then built by correlating the intensity
fluctuations obtained at different waist in the same simulation. The
biggest waist being 2/3 of the simulated matrix.

\begin{table}
\begin{tabular}{ll}
\hline \hline
 $T_{m,A}= 297.1 K$ & $\omega ^{gf}_{AA}=  1353 J.mol^{-1}$\\
 $T_{m,B}= 327.9 K$& $\omega ^{gf}_{BB}=  1474 J.mol^{-1}$\\
 $\Delta H_{A}= 13.165 kJ.mol^{-1}$& $\omega ^{gg}_{AB}=  607 J.mol^{-1}$\\
 $ \Delta H_{B}= 25.37 kJ.mol^{-1}$& $\omega ^{ff}_{AB}=  251 J.mol^{-1}$\\
 $\Delta S_{A}= 44.31 J.mol^{-1}.K^{-1}$& $\omega ^{gf}_{AB}=  1548 J.mol^{-1}$\\
 $ \Delta S_{B}= 77.36 J.mol^{-1}.K^{-1}$& $\omega ^{fg}_{AB}=  1716 J.mol^{-1}$\\
\hline \hline
\end{tabular}
\caption{\label{table1} \small{Monte-Carlo simulation parameters.}
\scriptsize{Parameters of the Monte Carlo simulation of DMPC-DSPC
mixture. The indices $g$ and $f$ correspond to gel and fluid state
respectively. Values are the one given in \cite{Hac05}}}
\end{table}

\subsection{Image analysis}
While running Monte-Carlo simulations, some images of the lattice
were recorded at a chosen Monte Carlo step frequency, using an
intensity code for the four different combination of the lipids
states ($I ^{AB}_{gf}$). This lead in easier numerical image
analysis capacities for thresholding.

Image correlation spectroscopy analyzes were conducted on
thresholded binary images using the ICS plugin for ImageJ
($http://rsb.info.nih.gov/ij/$) developed by Fitz Elliott, based on
the work of Petersen \cite{Pet93} which can be found at
$http://www.cellmigration.org/resource/imaging/software$. Using
spatial image correlation spectroscopy module of the plugin
generated a 2D correlogram of our images obtained from the MC
simulation. This 2D correlogram was reduced to a monodimensional
correlogram assuming a circular symmetry. The 1D correlogram was
arbitrary fitted by the following exponential function :

\begin{equation} \label{eqn:fitICS}
   g(\xi) = g(0)(a_{1}.e^{-(\xi/l_{c1})}+a_{2}.e^{-(\xi/l_{c2})})+g(\infty)
\end{equation}
where
% $l_{c1}$ is always found to be unity and
$l_{c_{2}}$ is the characteristic length of the domains.
Correlograms were average over 50 different images at different time
steps of the MC simulation at each temperature. The performance of
this procedure was tested using simulated images of circular domains
of different radius and showed error within 20 $\%$ accuracy.

Direct morphoanalysis was also performed using the ImageJ plugin
"particle analysis". Analysis are conducted on binary thresholded
image over a set of 50 different images obtained at different time
of the MC simulation as for ICS. The "particle anaylsis" plugin is a
simple binary derivative method that delimits areas of pixels having
the same value (1 or 0) therefore giving a direct access to the
number and area of the domains on each image. This method allows to
plot the exact distribution in term of area and occurrence of the
domains. For simplicity reason, it was decided to only keep the mean
area of the domains defined as the following :
\begin{equation} \label{eqn:sizedom}
  A_{dom}=\frac{1}{N}\sum_{A=2}^{A=\infty} {N_{A}}{A}
\end{equation}
with A the value of the area (discretised in classes every 25
pixels) and N the number of occurence of this area within the image.

Finally, the fractional area of gel lipids ($S_{g}^{n}$) defined as
the following $S_{g}^{n}=\frac{S_{g}}{S_{g}+S_{f}}$ within the image
could also be determined at each temperature using this method,
allowing therefore to scale the image in nm. This fraction was used
to extrapolate the total area of the simulation and the mean area
($\langle a \rangle=f_{g}.a_{g}+f_{l}.a_{l}$) occupied by a lipid
according the following value of lipid areas, $a_{f} = 63 {\AA}^{2}$
and $a_{g} = 45{\AA}^{2}$ for PC lipids \cite{Alm92,Wie89}.

%The area of the domains was simplified to be $\pi l_{c2}^2$ and could therefore be converted in terms of number of lipid content to be compared with values obtained by FCS laws.

\subsection{Preparation of multilamelar vesicles}

10 $\mu$l of a 10mM mixture of DMPC:DSPC (8:2 mol:mol) labeled with
$\beta$ Bodipy Fl C5-HPC at a lipid molar ratio of 1:50 000 were
deposited on a glass coverslip previously extensively cleaned with
ethanol, water:ethanol (7:3 vol:vol) and chloroform. Lipids were
dried under vacuum for one hour and hydrated with 700 $\mu$l of pure
distilled water previously heated at 50\r{°}C. Sample was then
allowed to cool down for at least 30 min to the different
measurement temperatures within the microscope chamber.

\subsection{FCS experiments}

FCS experiments were performed on a home-built device. This
experimental setup has been extensively described in \cite{Wen07} .
Briefly, it is based on an inverted microscope (Zeiss Axiovert 35M)
with a NA=1.2 water immersion objective (Zeiss C-Apochromat)and a
three-axis piezo-scanner (Physik Instrument, Germany). In order to
avoid photobleaching of the fluorescent label during the experiment
3$\mu$W of a CW laser at 488nm was used for fluorescence excitation.
For experiments above the diffraction limit, a 30$\mu$m pinhole
conjugated to the microscope object plane, defined the observation
volume. After the pinhole, the fluorescence signal is split by a
50/50 beamsplitter and focused on two avalanche photodiodes
(Perkin-Elmer SPCM-AQR-13) through a $535 \pm 20$~nm fluorescence
bandpass filters (Omega Filters 535DF40). The fluorescence intensity
fluctuations are analyzed by cross-correlating the signal of each
photodiode with a ALV6000 hardware correlator. For FCS diffusion
laws, waist of the laser in the object plane was modified by
underfilling the microscope objective back-aperture using an
adjustable diaphragm placed on the excitation optical path. The
values of the observation area therefore obtained were calibrated by
measuring the diffusion time of a 0.5 $\mu$M Rhodamine-6G in aqueous
solution at 22\r{°}C (D=280 $\mu m^{2}.s^{-1}$). For each waist at
each temperature of the study, at least 200 FCS measurements of 5s
duration each were made by series of 10.

\subsection{Fitting of ACFs}

For free Brownian two-dimensional diffusion in the case of a
Gaussian molecular detection efficiency and accounting for
negligible photophysics effects on the fluorophore (triplet state,
blinking, bleaching...) the fluorescence autocorrelation function
(ACF) is given by :

\begin{equation} \label{eqn:expACF}
   g^{(2)}(\tau) = 1+ \frac{1}{N}\frac{1}{1+\frac{\tau}{\tau_d}}
\end{equation}

where N denotes the average number of molecules in the observation
area and $\tau_{d}$ the diffusion time.$\tau_{d}$ is linked to the
laser beam transversal waist w and the molecular diffusion
coefficient D by :
\begin{equation} \label{eqn:diff}
   \tau_{d}=\frac {w^{2}}{4D}
\end{equation}

All the ACFs obtained experimentally using variable waist above
diffraction limit, or on Monte-Carlo simulation were fitted using
Eq.~\ref{eqn:expACF}.

\section{Results}

\subsection{FCS at variable radii and variable temperature on DMPC and DMPC:\\DSPC
samples}

\begin{figure}[!h]
   \begin{center}
      \includegraphics*[width=4.5in,height=4.5in]{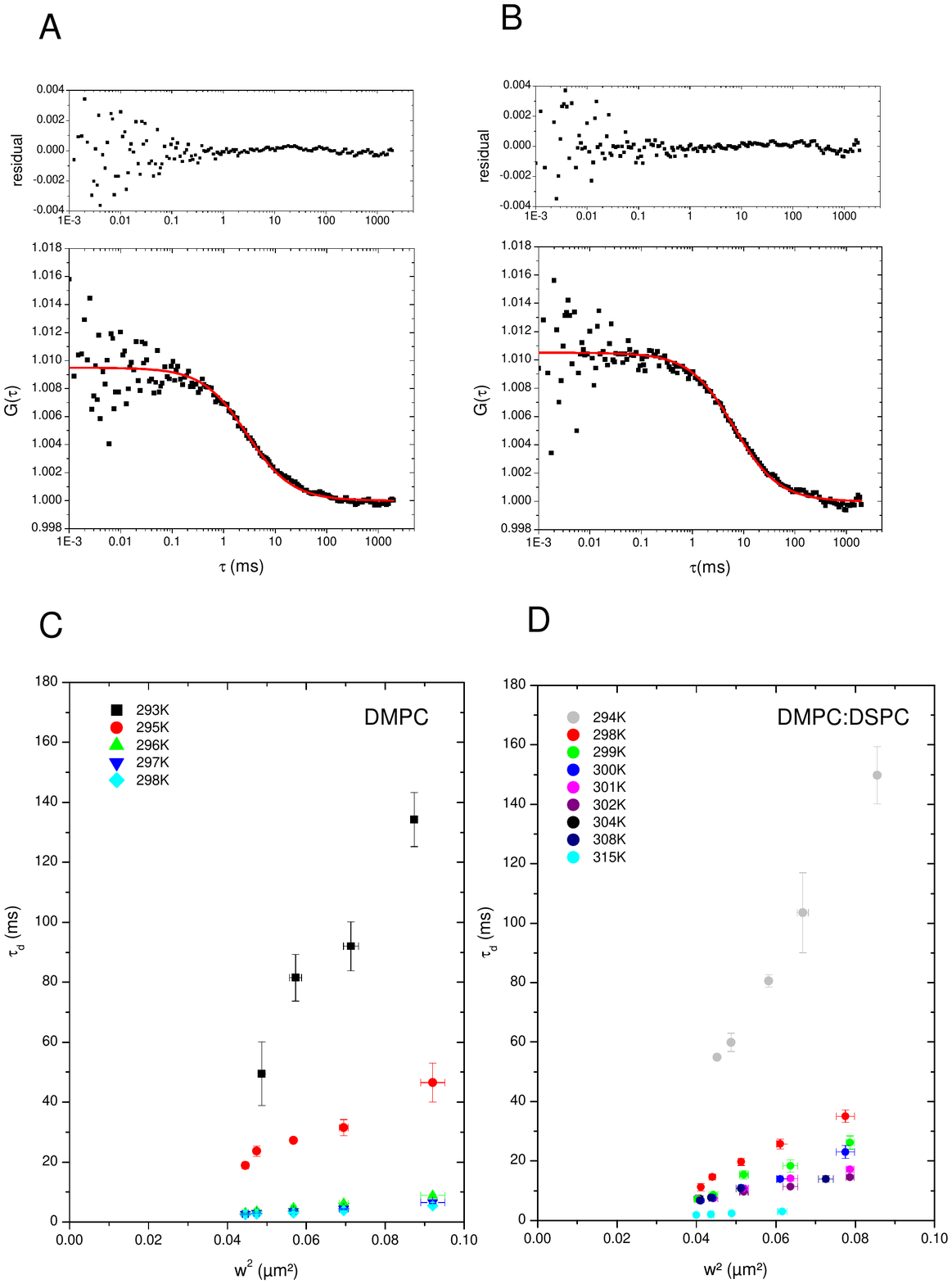}
      \caption{\small{\textbf{Experimental correlograms and FCS
diffusion laws obtained on lipid mixtures.}}
 \scriptsize{\textbf{Part A and B}
Correlograms obtained on multilamellar vesicles of DMPC (Part A) and
DMPC:DSPC 8:2 mol:mol (Part B) (square dots are experimental data)
fitted using eq.~\ref{eqn:expACF}. Residuals are plotted in the
upper part of the figure showing accuracy of the fit. \textbf{Part C
and D} Experimental FCS diffusion laws obtained for DMPC (Part C)
and DMPC:DSPC 8:2 mol:mol (Part D) at different temperatures.
(Colors on line)}}
      \label{fig2}
   \end{center}
\end{figure}

FCS diffusion laws were acquired on pure DMPC lipid multi-lamellar
vesicles and on vesicles made of DMPC:DSPC (8:2 mol:mol) both
labeled with C5-Bodipy-PC at different temperatures below, within
and above the phase transitions. Fig~\ref{fig2} A shows a typical
correlogram obtained on pure DMPC at 297K with a transversal waist
$w$=218 $\mu$m, Fig~\ref{fig2} B shows the same type of experimental
correlogram obtained on the DMPC:DSPC 8:2 mol:mol mixture at 302K
with a transversal waist $w$=210 $\mu$m. On both part of the figure,
the red line is the fit of the correlogram using
Eq.~\ref{eqn:expACF} allowing determination of $\tau_{d}$ in the
correlogram. The residual, which is plotted in the upper part of Fig
~\ref{fig2} A and B, confirms the quality of the fit. Repeating
experiments at different waists and different temperatures allows to
obtain FCS diffusion laws. These are depicted in Fig ~\ref{fig2} C
and D for pure DMPC and DMPC:DSPC mixture respectively.

FCS diffusion laws can be fit at different temperatures using the
following equation :
\begin{equation} \label{eqn:linfit}
   \tau_{d}= \frac{w^2}{4D_{eff}}+\tau_{d_{0}}
\end{equation}
with $\tau_{d}$ the experimental values of diffusion time at
different waist $w$ and $\tau_{d_{0}}$ the extrapolated diffusion
time at $w^2=0$. These FCS diffusion laws also allow the
determination of an effective diffusion coefficient named $D_{eff}$.
Fig.~\ref{fig1s} A and B at the end of this paper show the increase
of this diffusion coefficient $D_{eff}$ with temperature for both
sample.
%The experimentally obtained results for this D show a very
%high similarity to the one already found by Vaz \textit{et al.},
%\cite{Vaz89} by FRAP experiments in DMPC:DSPC 80:20 mol:mol
%mixtures (\emph{trouver ref pour DMPC seul}).

\begin{figure}[!h]
   \begin{center}
      \includegraphics*[width=4.5in,height=4.5in]{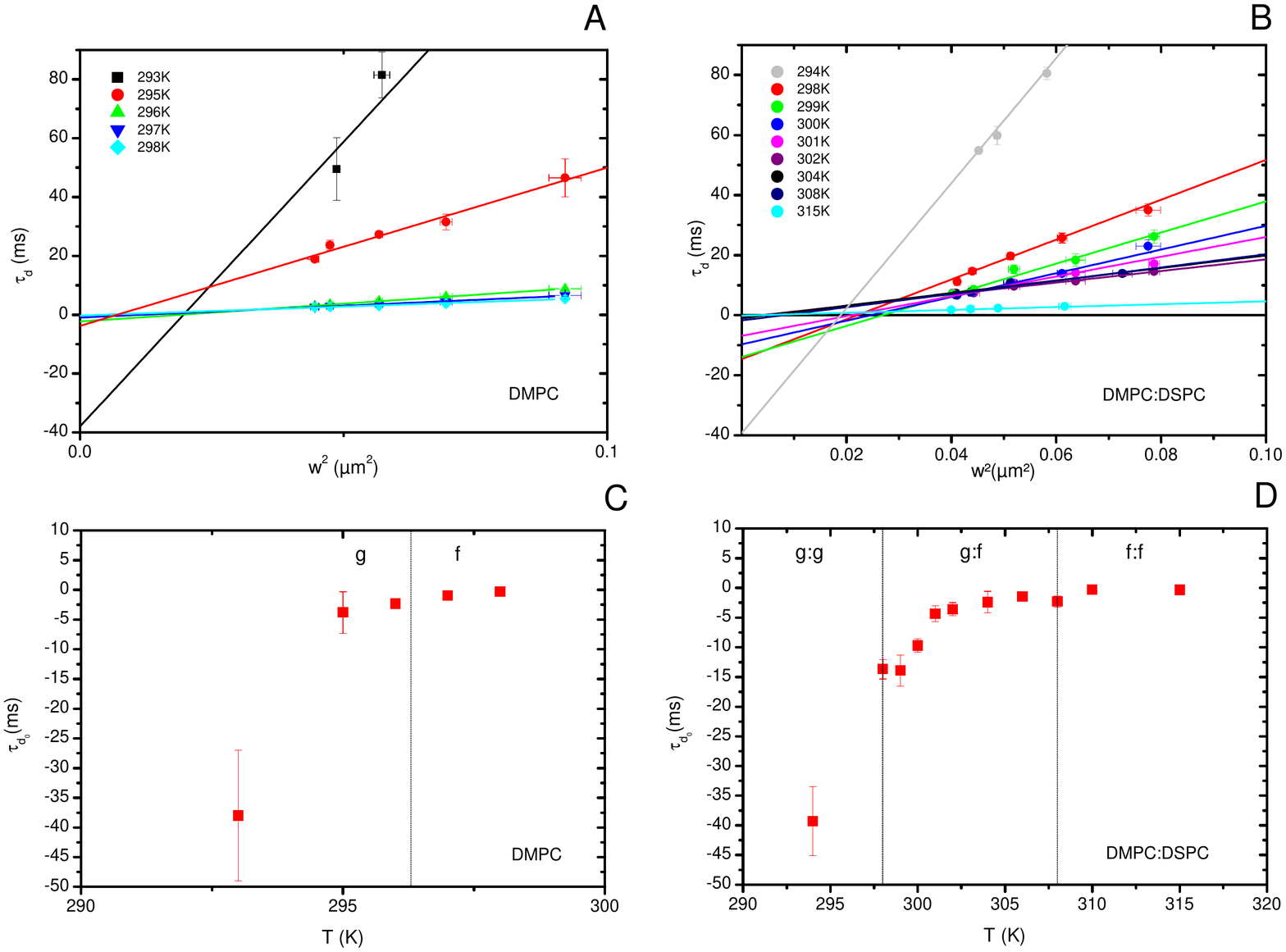}
      \caption{
      \small{\textbf{Variation of experimental
$\tau_{d_{0}}$ as a function of temperature.}}
\scriptsize{\textbf{Part A and B} Fit of the experimental FCS
diffusion laws using eq.~\ref{eqn:linfit} for DMPC (Part A) and
DMPC:DSPC 8:2 mol:mol (Part B) allowing determination of
$\tau_{d_{0}}$ for each set of data. \textbf{Part C and D}
$\tau_{d_{0}}$ is plotted against temperature for DMPC (part C) and
DMPC:DSPC 8:2 mol:mol (part D). Square dots are obtained value with
error bar. (Colors on line)}}
      \label{fig3}
   \end{center}
\end{figure}

As illustrated in Fig~\ref{fig3} A and B, fit of the experimental
FCS diffusion laws exhibit a negative $\tau_{d_{0}}$ for
temperatures below the phase transition (Fig ~\ref{fig3} A,
$T<297K$) in the case of DMPC or below the second transition
($gf:ff$) in the case of the DMPC:DSPC mixture (Fig ~\ref{fig3} B,
$T<310K$). When the FCS laws are acquired in pure fluid phase
($T>310K$ for DMPC:DSPC mixture or $T>298K$ for DMPC alone), they
exhibit $\tau_{d_{0}}$ values close to zero which shows a pure
Brownian behavior of the diffusing molecules as expected. When
plotting these $\tau_{d_{0}}$ value as a function of temperature, it
can be seen on Fig ~\ref{fig3} C and D that in both cases (DMPC
alone or lipid mixture) the more the system is in the gel phase (or
the more the temperature decrease) the more $\tau_{d_{0}}$ becomes
negative.

In order to confirm the experimental results and to understand the
origin of a negative $\tau_{d_{0}}$ when gel domains are present in
a fluid phase, it was decided to perform Monte Carlo simulations
using a full set of thermodynamic parameters for DMPC and DMPC:DSPC
lipid mixtures.

%Since variation of D and $\tau_{d_{0}}$ observed experimentally by FCS diffusion laws seemed to be strikingly identical to the one obtained with MC simulation, it was decided to analyze the variation of experimental $w^{2}_{0}$ as a function of temperature. $w^{2}_{0}$ was calculated from the experimental FCS diffusion laws as previously stated.

%In Fig~\ref{fig9} are shown the values obtained for DMPC:DSPC 80:20 mixture (Fig~\ref{fig9}A) and DMPC alone (Fig~\ref{fig9}B). Here again, these curves exhibit two maxima in the case of the DMPC:DSPC 80:20 mixture at 299K and 306K and one in the case of DMPC alone at 296K, these value being close to the respectively $gg:gf$ and $gf:ff$ transition for the lipid mixture $g:f$ transition with the lipid alone.
%For the lipid mixture, $w_{0}^{2}$ values are in between 10 000 and 26 000 nm$^{2}$ which give value of $w_{0}$ higher than 50 nm and lower than 80 nm.

\subsection{FCS at variable radii on MC numerical simulations : Diffusion
Laws}

\begin{figure}
   \begin{center}
      \includegraphics*[width=3.5in,height=4in]{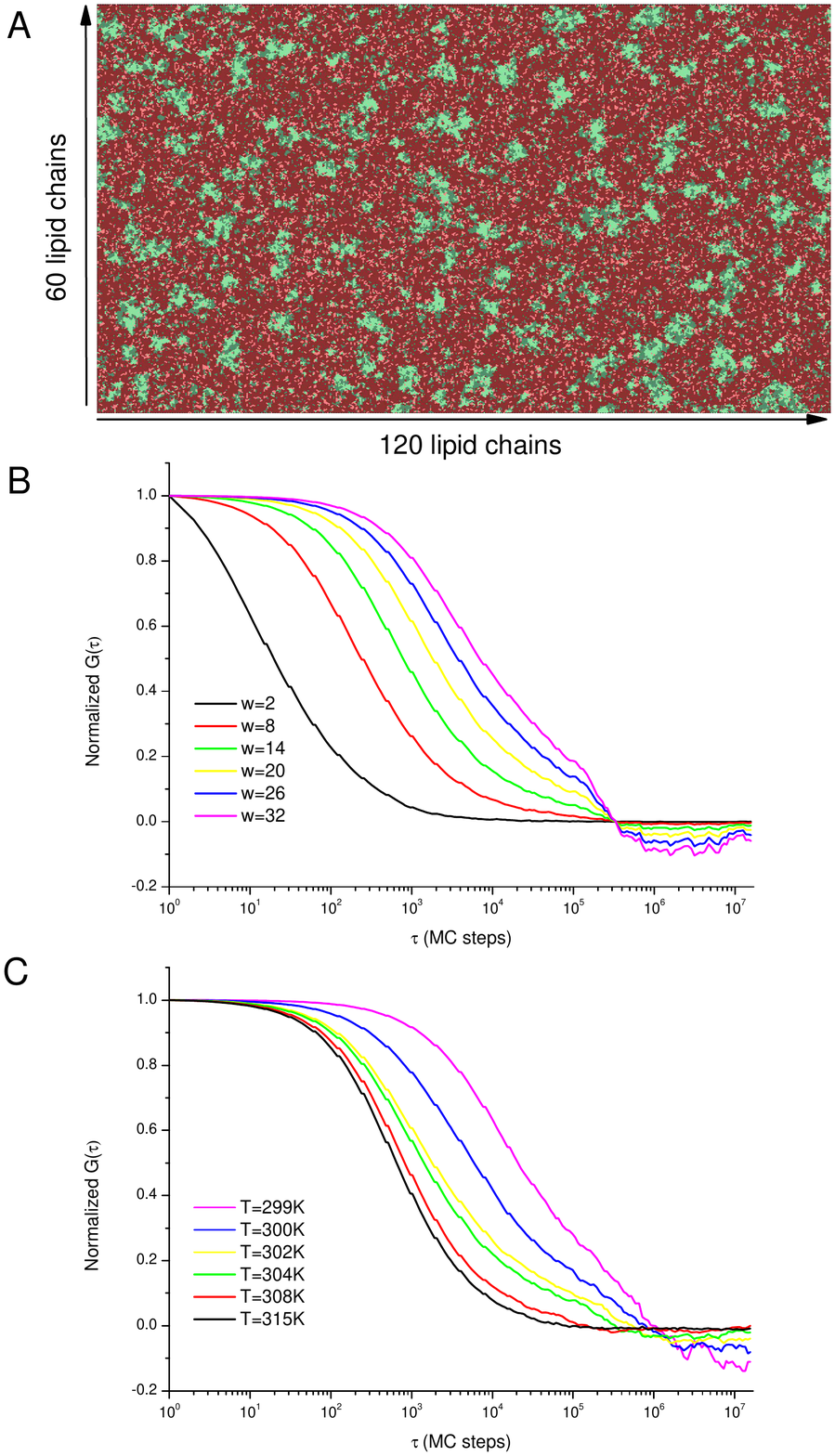}
      \caption{\small{\textbf{Images and Correlograms obtained by
MC simulation of DMPC-DSPC 8:2 mol:mol mixture.}}
\scriptsize{\textbf{Part A}. Monte-Carlo snapshot of a 60x60 lipids
lattice 8:2 mol:mol DMPC-DSPC mixture at 304K. Green domains
correspond to gel lipids, red domains to fluid lipids. \textbf{Part
B}. Normalized correlograms obtained from the Monte-Carlo simulation
at 304K for different waists. \textbf{Part C}. Normalized
correlogram obtained from the Monte-Carlo simulation at a given
waist (w=8 l.u.) at different temperatures from gel ($T<300K$) to
liquid ($T>315K$) and within the melting regime ($300K<T<315K$).}}
      \label{fig4}
   \end{center}

\end{figure}

Monte Carlo simulation were made at different temperature below,
within and above the two state transitions $(gg:gf:ff)$ of the
DMPC:DSPC 8:2 (mol:mol) and below, within and above phase transition
of DMPC alone ($g:f$). Fig~\ref{fig4}A shows a snapshot of the
DMPC:DSPC 8:2 mol:mol mixture at 304K ($gf$ state) as obtained from
the MC simulation. Gel domains (in green) can be seen in a sea of
fluid lipids (in red)(60x120 lipid chains). MC simulated
correlograms of this mixture are depicted for different waist at
this temperature in Fig~\ref{fig4}B, showing an increase of
$\tau_{d}$ with increasing waists as expected.  As seen from
experiments (see Fig~\ref{fig3}), $\tau_{d}$ is also expected to
increase with decreasing temperature at a given waist, this is
illustrated in Fig~\ref{fig4}C, confirming the validity of our MC
simulations. At low temperature ($T<300K$) or large waist ($w>20
l.u.$) the correlation function at long time scale shows higher
noise due to the limited number of events arising at these times.
Nevertheless, the determination of $\tau_{d}$ is still valid.
%As it can be seen from Fig.~\ref{fig3}A $\tau_{d}$ is increasing with lowering the temperature as it could be expected. Fig.~\ref{fig3}B shows the correlograms obtained for different waist at 299K (close to the first transition). From this figure it can be seen that $\tau_{d}$ is increasing with increasing radius also as expected.
FCS diffusion laws could therefore be established here again by
plotting $\tau_{d}$ as a function of the square of the waist $w^2$.

\begin{figure}
   \begin{center}
      \includegraphics*[width=3in,height=4in]{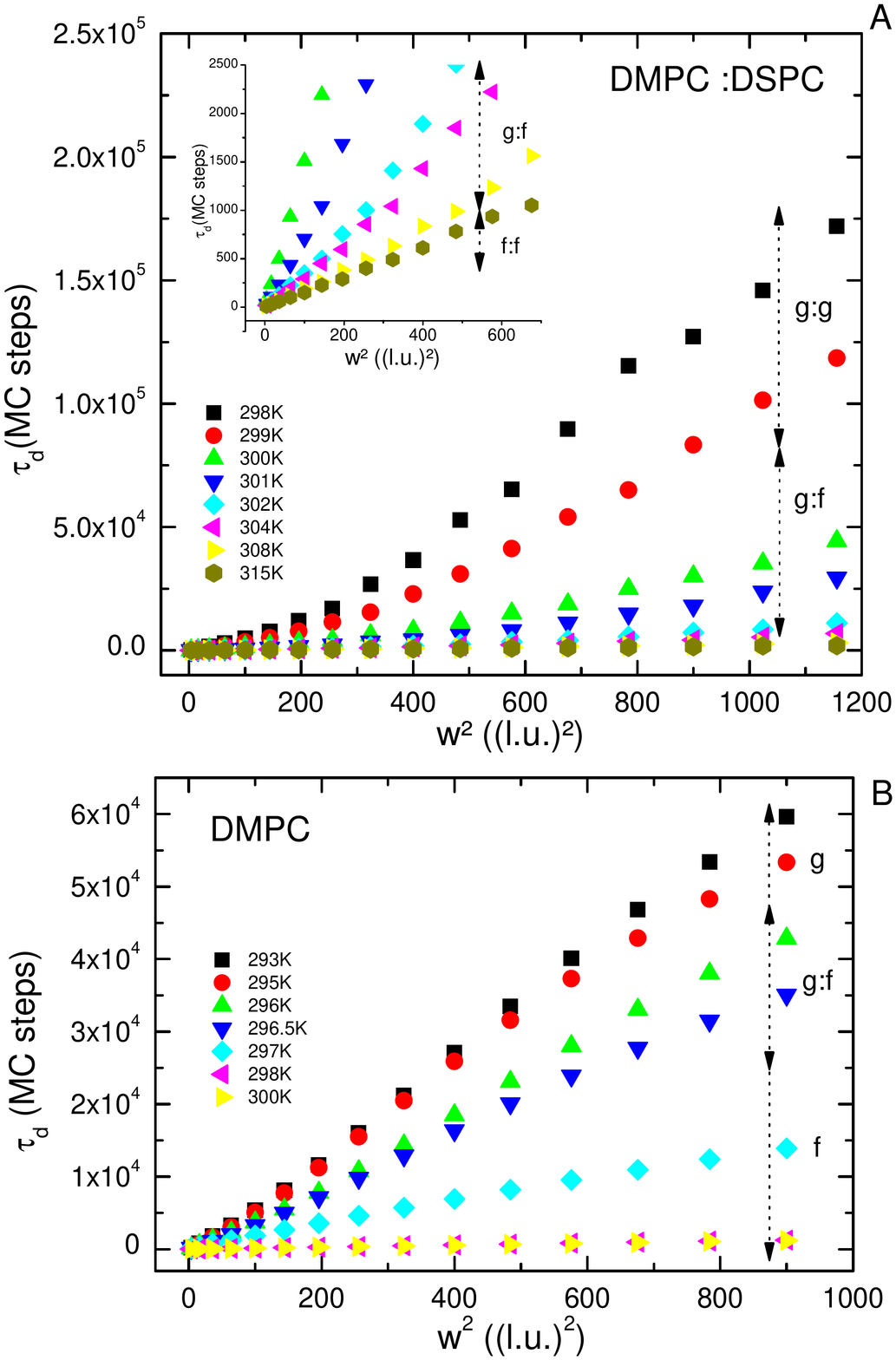}
      \caption{\small{\textbf{FCS diffusion laws obtained from MC
simulation at different temperatures.}} \scriptsize{ \textbf{Part A}
FCS diffusion laws obtained from MC simulated DMPC-DSPC 8:2 mol:mol
mixture are represented for different temperature below, within and
above the melting regime of the mixture. Onset shows an enlargement
for low waists at high temperature. These FCS diffusion laws exhibit
two slopes at low temperature but starts to be linear with a null
origin at temperature above 310K, indicating a pure Brownian
diffusion process. \textbf{Part B}  FCS diffusion laws obtained from
MC simulations of pure DMPC are represented for different
temperature below and above phase transition (296.3K). FCS diffusion
laws exhibit here again two slopes below the phase transition and
starts to be linear with a null origin at temperature above 298K.}}
      \label{fig5}
   \end{center}
\end{figure}

As illustrated in Fig.~\ref{fig5} FCS diffusion laws are obtained
for the different temperature of the MC simulations for both
DMPC:DSPC mixture (Fig.~\ref{fig5}A) and DMPC alone
(Fig.~\ref{fig5}B). As found experimentally in the case of DMPC:DSPC
mixture, above 310 K the FCS laws seems perfectly linear (Fig
~\ref{fig5} A), accounting for a pure Brownian diffusion of the
lipids. Below 310 K (Fig ~\ref{fig5} A inset) all the FCS laws seems
to be biphasic, exhibiting therefore deviations from a pure Brownian
diffusion as expected for inhomogeneous media. Even below the first
transition temperature gg : gl of the lipid mixture these FCS laws
still shows a biphasic behavior. As shown in fig.~\ref{fig5} B the
same biphasic type of FCS diffusion laws has been obtained with pure
DMPC MC simulations below the transition temperature (296.5 K).

\begin{figure}
   \begin{center}
      \includegraphics*[width=4.5in,height=3.5in]{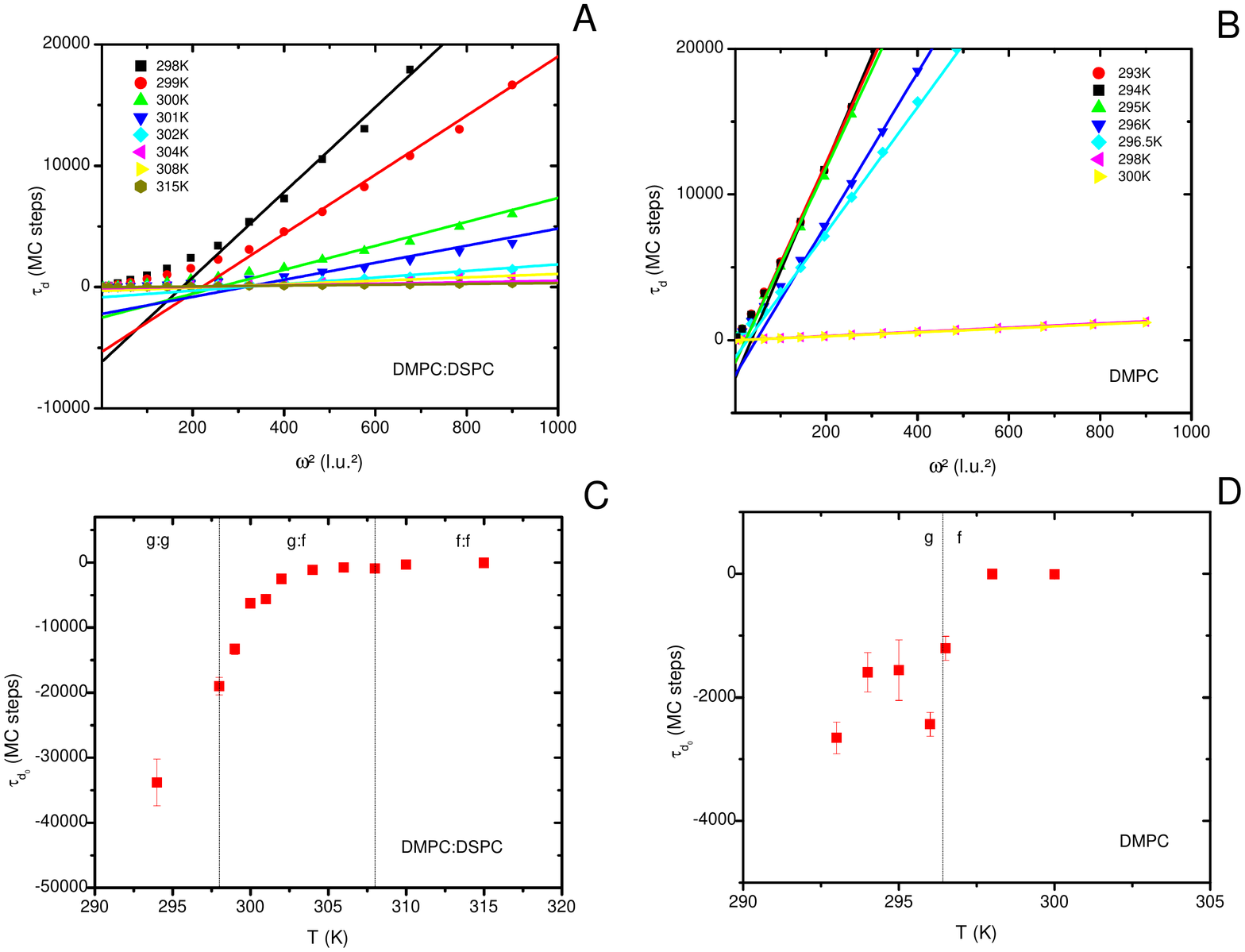}
      \caption{\small{\textbf{Variation of $\tau_{d_{0}}$
obtained by simulations as a function of
temperature.}}\scriptsize{\textbf{Part A and B}.  Fit of the
asymptotic part of the FCS diffusion law using eq~\ref{eqn:linfit}
at different temperatures for DMPC:DSPC 8:2 mol:mol (part A) and
DMPC (part B) \textbf{Part C and D}. $\tau_{d_{0}}$ is plotted
against temperature for DMPC:DSPC 8:2 mol:mol (part C) and DMPC
(part D). Square dots are obtained value with error bar. A sharp
increase is seen around 300K, close to the first meting temperature
of the DMPC-DSPC 8:2 mol:mol mixture (part C) and around 296K, close
to the melting temperature for DMPC.}}
      \label{fig6}
   \end{center}
\end{figure}

MC simulated FCS diffusion laws can be fit using the equation
~\ref{eqn:linfit} as depicted in Fig.~\ref{fig6} A for DMPC:DSPC
mixture and in Fig.~\ref{fig6} B for DMPC alone.

As in the experiments, the value of an effective diffusion
coefficient $D_{eff}$ can be calculated following
Eq.~\ref{eqn:linfit} and is plotted as a function of temperature in
supplementary material Fig~\ref{fig1s} C (DMPC:DSPC mixture) and D
(DMPC alone). In both case, it is important to note that the global
shape of the curve $D_{eff}=f(T)$ is the same for simulation and
experiments and is signed by a sharp increase during the transition
from gel to fluid medium.

It can be seen that, as in the experimental FCS diffusion laws,
$\tau_{d_{0}}$ decrease with decreasing temperature as plotted in
Fig.~\ref{fig6}C and D respectively for DMPC:DSPC mixture and DMPC
alone. These $\tau_{d_{0}}$ also reach values close to zero at
$T\geq 310 K$ for the DMPC:DSPC mixture or $T\geq 298 K$ for DMPC
alone, confirming a pure Brownian diffusion of the system above this
temperature.

These observations confirm the validity of our experimental results
on both DMPC and DMPC:DSPC lipid mixture. Both system exhibit
deviation from a pure Brownian motion as soon as they are not in
pure fluid state. These deviations can be seen using FCS diffusion
laws and are signed by a negative $\tau_{d_{0}}$.

%Surprisingly, $\tau_{d_{0}}$ values were all found to be negative at temperature below the second transition ($gf:ff$). This will be discussed later in term of relative diffusion and confinement time in the heterogeneities of the lipid lattice.

\subsection{Origin of negative diffusion time at zero waist}

As shown above in Fig~\ref{fig3}C and D or Fig~\ref{fig6}C and D,
$\tau_{0}$ extrapolated from the fit of the FCS diffusion laws (both
experimental and simulated) using eq ~\ref{eqn:linfit} exhibit
growing negative value with decreasing temperature. This result is
surprising since it is different from the one obtained by
Wawrezinieck \textit{et al} \cite{Waw05} in the case of microdomains
(where $\tau_{d_{0}}$ is positive) since DMPC/DSPC mixture domains
would have been expected to behave similarly. The results obtained
here are closer to the one obtained in this previous study for
meshworks (see Fig~\ref{fig1}for comparison). In this later case,
Wawrezinieck \textit{et al} \cite{Waw05} shown that the FCS
diffusion law could be correctly described by the following
equations :

\begin{eqnarray} \label{eqn:linfitmesh}
   \tau_{d}=
    \begin{cases}
    \frac{w^2}{4D_{micro}}  & \text {if $X^{2}_{c}<2$} \\
             S_{conf}\frac{w^2}{4D_{micro}}+k(\tau^{domain}_{d}-\tau_{conf}) &\text {if $X^{2}_{c}>2$}
    \end{cases}
\end{eqnarray}
Where $D_{micro}$ is the local diffusion coefficient,
$\tau^{domain}_{d}$ is the characteristic diffusion time inside a
mesh, $\tau_{conf}$ is the confinement time defined as
$S_{conf}\tau^{domain}_{d}$ and $S_{conf}$ is the strength
confinement that is a function of the probability to cross a barrier
to go from one mesh to another, giving therefore an indication on
the stiffness of the barrier. $X^{2}_{c}$ being the normalized
coordinates of the waist($w$) to the meshsize ($r$) and defined in
the case of a square meshwork as $X^{2}_{c}=\frac{\pi
w^{2}}{4r^{2}}$. It has to be noted that for $X^{2}_{c}>2$
eq~\ref{eqn:linfitmesh} can be rewritten as :
\begin{equation} \label{eqn:linfitmeshref}
 \tau_{d}=S_{conf}\frac{w^2}{4D_{micro}}+k\tau^{domain}_{d}(1-S_{conf})
\end{equation}
%Which shows that for $S_{conf}=0 \tau_{d}=\tau^{domain}_{d}$.
%This set of equations shows that the FCS diffusion law could be described by two linear laws crossing over close to the mesh size, which authorize measurement of it.

In the case of the DMPC/DSPC mixture, one can simplify the system to
a two state (gel and fluid) system with for each a typical diffusion
time $\tau^{g}_{d}$ and $\tau^{f}_{d}$ exist. It has to be noted
that at least for the MC simulation, at a given waist,
$\tau^{g}_{d}= 70 \tau^{f}_{d}$. Since $\tau^{g}_{d}$ is much bigger
than $\tau^{f}_{d}$, when a tracer moves into a gel environment it
can appear as being strongly confined to this environment as
compared to the same tracer into a fluid environment. Therefore, gel
domains can be considered as barriers to free diffusion of the
molecule in the liquid phase.
%as
%this can be seen from the movie presented in the supplementary
%material.
It can be clearly seen from a movie generated using images step by
step in the simulation (data not shown) that the tracer molecules
move freely in the fluid environment while they seem to be unable to
cross a gel environment at least during the duration of the movie
(300 MC steps).
% this later acting as a barrier
%for diffusion of the molecule.
Therefore, by analogy to the meshwork case described here above,
$\tau_{conf}$ could be defined as a function of $\tau^{g}_{d}$ times
the normalized surface of gel ($S_{g}^{n}$) that has to be crossed
by the tracer molecule.

FCS diffusion law obtained here could be described by the following
simple intuitive model :

\begin{eqnarray} \label{eqn:linfitlip}
  \tau_{d}=
    \begin{cases}
    \frac{w^2}{4D_{micro}(T)}   & \text {if $w^{2}<aw^2_{0}$}\\
    S_{g}^{n}\frac{w^2}{4D_{micro}(T)}+K(\tau^{f}_{d}-S_{g}^{n}\tau^{g}_{d}) & \text {if $w^{2}>aw^2_{0}$}
    \end{cases}
\end{eqnarray}
With $K$ and $a$ given constants and $w^2_{0}$ the value of $w^2$
for $\tau_{d}=0$.

Eq ~\ref{eqn:linfitlip} shows that the first part of the the FCS
diffusion law, below the cross over regime, is linked to the local
diffusion constant $D_{micro}$ which is a function of the
temperature. In the MC simulation, $D_{micro}$ has been defined such
that $D_{micro}^{f}=70D_{micro}^{g}$. By linear fit of the simulated
FCS laws at small waists ($w^{2}< 300 l.u. ^{2}$), $D_{micro}$ can
be found for each temperature and it can be found that in the case
of MC simulation $D_{micro}(325K)= 70D_{micro}(294K)$ (data not
shown). If one extrapolate the value of D from the second part of
the FCS diffusion law ($300 l.u. ^{2}<w^{2}< 1200 l.u. ^{2}$), then
it is found that $D_{micro}(325K)= 160 D_{micro}(294K)$. This
confirm that the shape of second part of the FCS diffusion law does
contain more than only free diffusion. This cannot be verified for
experimental laws since the first part of the FCS law is not
accessible.

As defined above, $\tau_{d_{0}}$ is the value of $\tau_{d}$ for
$w^2=0$. From eq~\ref{eqn:linfitlip} it is found that :

\begin{equation} \label{eqn:taudzero}
   \tau_{d_{0}}=K\tau^{f}_{d}(1-70S_{g}^{n})
\end{equation}

\begin{figure}
   \begin{center}
      \includegraphics*[width=3in,height=4in]{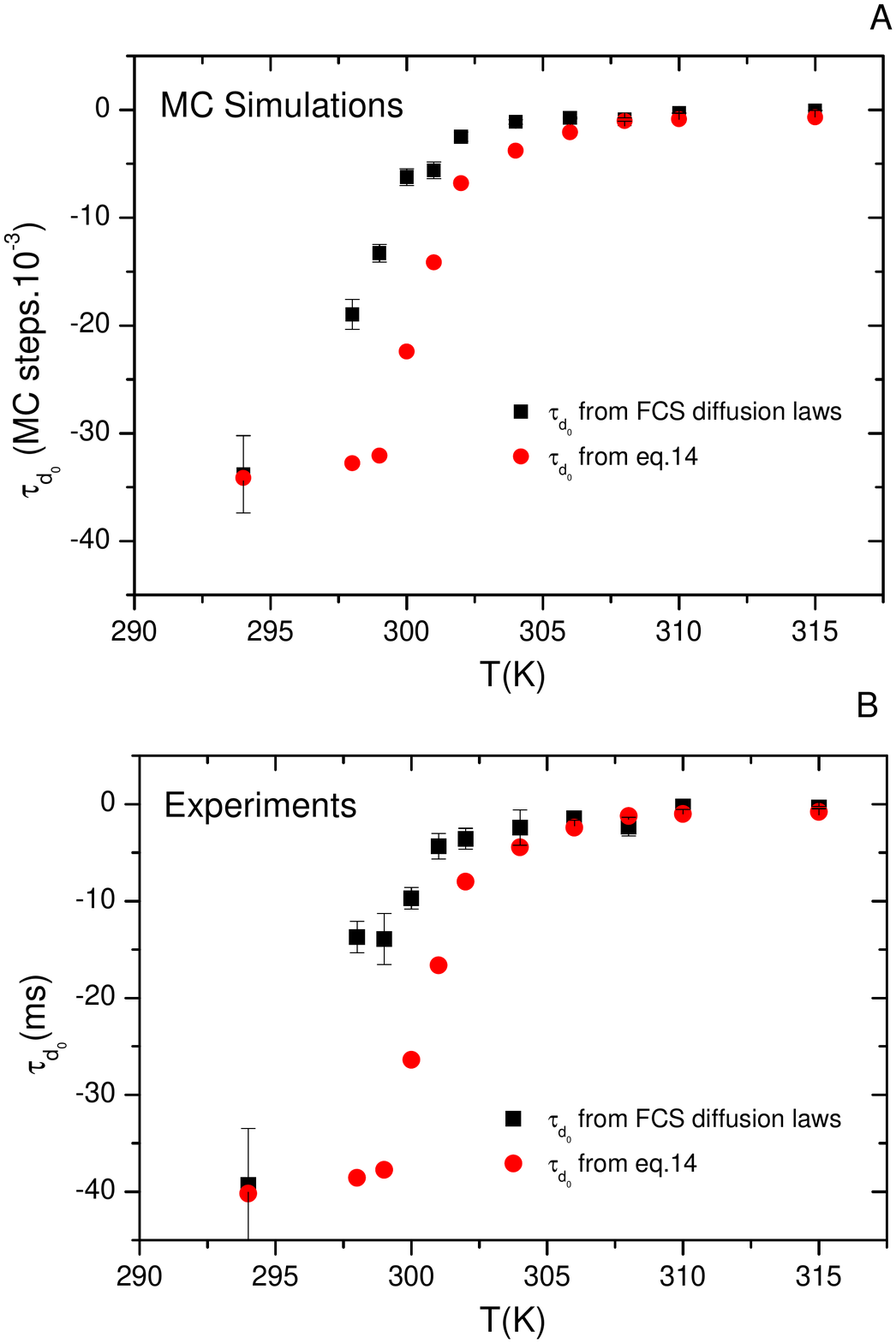}
      \caption{\small{\textbf{Comparison of experimental and MC
simulated $\tau_{d_{0}}$ to $\tau_{d_{0}}$ from
eq.~\ref{eqn:taudzero}.}} \scriptsize{\textbf{Part A}. Comparison of
$\tau_{d_{0}}$ obtained by linear fit of the asymptotic part of the
simulated FCS diffusion laws to $\tau_{d_{0}}$ obtained by
eq.~\ref{eqn:taudzero} as a function of temperature ( i.e. a
function of normalized gel area). \textbf{Part B}. Comparison of
$\tau_{d_{0}}$ obtained by linear fit of the asymptotic part of the
experimetal FCS diffusion laws to $\tau_{d_{0}}$ obtained by
eq.~\ref{eqn:taudzero} as a function of temperature. Both are
normalized by means of K factor at 294K.}}
      \label{fig7}
   \end{center}
\end{figure}

With increasing temperatures, the relative normalized solid area
($0<S_{g}^{n}<1$) is decreasing down to zero. In this study,
$S_{g}^{n}$ was determined from MC simulation of the DMPC/DSPC
mixture at different temperatures. Figure~\ref{fig7} shows evolution
of $\tau_{d_{0}}$ with temperature for MC simulated FCS diffusion
laws (Fig.~\ref{fig7} A) and experimental FCS diffusion law
(Fig.~\ref{fig7} B) and the calculated values of $\tau_{d_{0}}$ as a
function of $S_{g}^{n}$ according to eq.~\ref{eqn:taudzero},
adjusted to the lowest temperature value of $\tau_{d_{0}}$ by K
constant. Both figures ~\ref{fig7} A and B shows a nice adequation
of the calculated values with eq.~\ref{eqn:taudzero} and the
numerical or experimental obtained values confirming the hypothesis
that solid domains act as barriers to free-diffusion of the tracers.
Although the present model is rather heuristic than accurate, it
suggests that the gel confinement areas can be viewed as meshwork
barriers for the diffusion of fluorescent reporters, but not as
impermeable domains, or as raft-like domains.

%In this simple model, only solid and liquid domains has been considered as the component of the lipid mixture therefore leading to two different types of diffusion characteristic times depending on the partition of the tracer. But fluctuations times of the mixture has also to be taken into considerations. Besides of these two diffusion characteristic times ($\tau^{g}_{d}$ and $\tau^{f}_{d}$), a fluctuation characteristic time ($\tau_{fluc}$) accounts for the frequency of one lipid to change its state from liquid to solid. It is known that this time goes through a maximum at the transition temperature \cite{Wun09,Gra02}. From figure ~\ref{fig10} it can be seen that there is a deviation from the measured $\tau_{d_{0}}$ and the estimated $\tau_{d_{0}}$ from eq ~\ref{eqn:taudzero} which starts from 298 to 308 K with a maximum at 299K in the case of MC simulations FCS diffusion laws and 298K in the case of experimental FCS diffusion laws. These values are close to the first transition temperature ($gg:gl$) estimated at 300K. It is clear that this deviation is linked to the fluctuation time and therefore the FCS diffusion law could be rewritten the following way :

%\begin{eqnarray} \label{eqn:linfitlip2}
%   \tau_{d}=
%    \begin{cases}
%    \frac{w^2}{4D_{micro}(T)}  & \text {if $w^{2}<aw^2_{0}$}\\ \\
%    S_{g}\frac{w^2}{4D_{micro}(T)}+K(\tau^{l}_{d}-S_{g}\tau^{g}_{d})+\tau_{fluc}& \text {if $w^{2}>aw^2_{0}$}\\
%    \end{cases}
%\end{eqnarray}

%This lead to a maximum value of $\tau_{fluc}$ in multilamellar vesicle close to 25 ms at 298K.

\subsection{Determination of domain size}

Wawrezinieck \textit{et al} \cite{Waw05} have previously shown that
deviation from a pure linear regime in the FCS diffusion laws were a
signature of heterogeneities in the probed environment. They also
showed that the crossover regime in the FCS diffusion laws was a
function of the relative size of both the probing waist and the
domain size. In the case of mesh restriction to diffusion,
Eq.~\ref{eqn:linfitmesh} shows that the FCS diffusion law could be
described by two linear laws crossing over each other close to the
mesh size, therefore authorizing measurement of it. The FCS
diffusion laws obtained from MC simulations in this study can also
be described by two linear laws crossing over at a given waist,
exhibiting a cross over regime which is a putative function of the
domain size.

These MC simulated FCS laws were analysed by a simple linear fit of
the highest slope of the curves (from $300 l.u. ^{2}<w^{2}< 1200
l.u. ^{2}$), using equation ~\ref{eqn:linfit}, since the first part
of the curve is not accessible experimentally and therefore the
cross over regime cannot be directly determined. Nevertheless,
eq~\ref{eqn:linfitlip} shows that the cross-over regime should occur
at $aw^{2}_{0}$. Since $w^{2}_{0}$ is the parameter that can be
easily determined from both simulations and experiments, it seemed
important to estimate in which extent it is related to the size of
domains.

For this purpose, snapshots of the simulations as the one depicted
in Fig ~\ref{fig4} A were made at different temperatures for the
DMPC:DSPC lipid mixture. These images were analyzed by ICS and by
direct space morphoanalysis as described previously.

%The main advantage of ICS is to analyze fluctuations in space within the image, giving therefore a mean typical size of distinguishable objects that can be considered as a coherence length ($l_{c}$). This will not be easy to interpret by any other type of image treatment.
In the case of ICS analysis, all images between the two main phase
transitions were analyzed using a biexponential function with two
characteristics ($l_{c_{1}}$ and $l_{c_{2}}$) coherence lengths. It
appeared that, in that case, $l_{c_{1}}$ varied between 0.5 and 1
$(l.u.)^2$ which is a value equivalent to one chain or one lipid and
indeed the smallest motif the image can contain. Whereas for the
direct space morphology analysis, domain representing less than two
lipids were not taken into consideration. Comparison of values found
for each methods are made by measuring the number of lipids in the
domains. This is directly obtained on the morphoanalysis and
estimated by FCS and ICS in supposed circular domains of area $\pi
l_{c_{2}}^{2}$ or $\pi\frac{\sqrt {w^{2}_{0}}}{2}$. There are no
fundamental reasons for choosing circular domains except
simplification for the comparison of each analysis. Indeed the
snapshot of the simulations clearly shows that the domains are not
circular nor square but have a very complex geometry, which has been
already shown in detail by Sugar \textit{et al.} \cite{Sug01}.

\begin{figure}
   \begin{center}
      \includegraphics*[width=4in,height=3in]{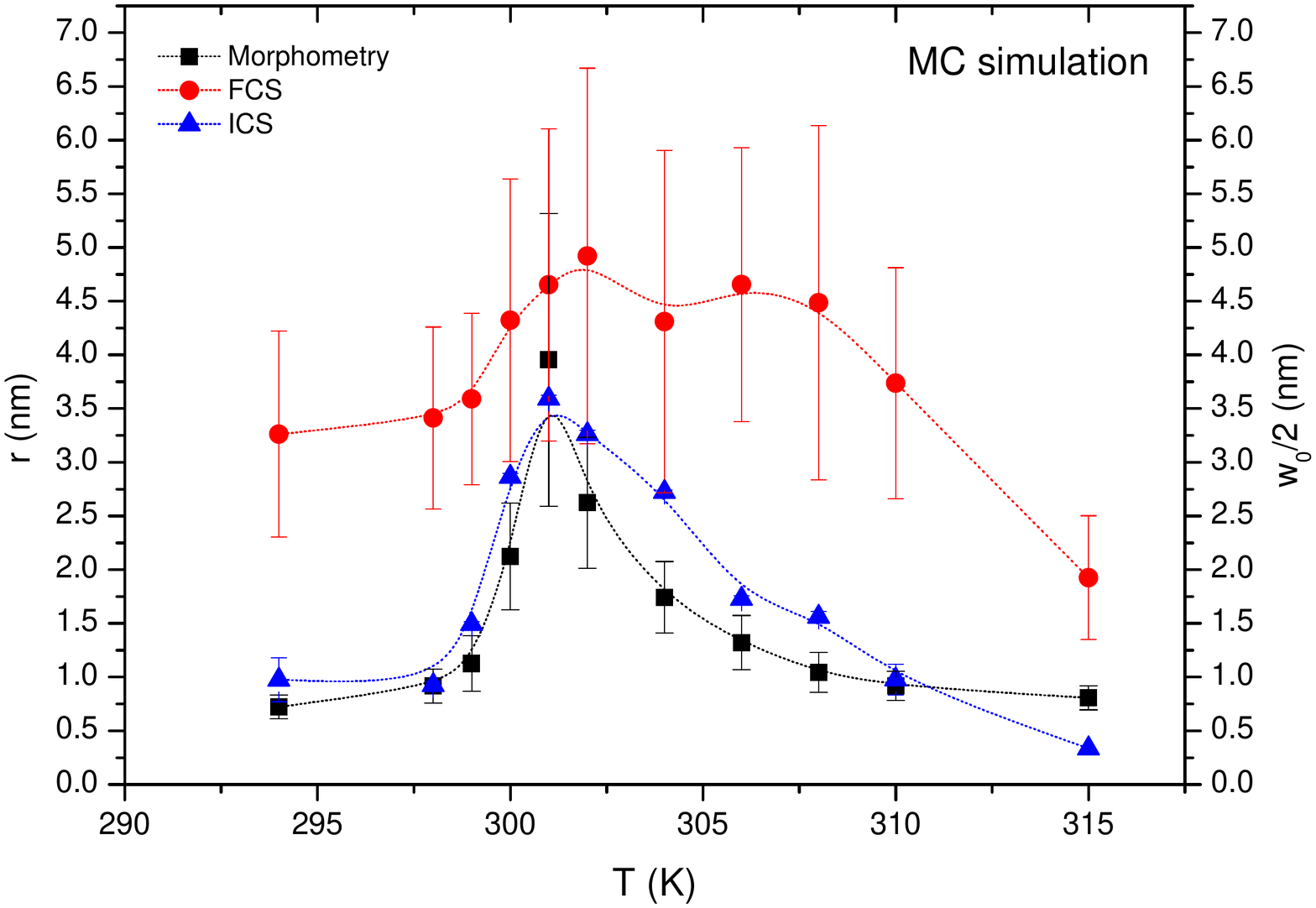}
      \caption{\small{\textbf{Comparison of $w^{2}_{0}$ obtained
by three different methods as a function of temperature.}}
\scriptsize{Three different methods (see text for details) were used
to determine the mean size of the domains within the MC simulation
of DMPC:DSPC 8:2 mol:mol mixture. These methods are compared as a
function of temperature. Circles are the values obtained by
determination of $w_{0}^{2}$ in FCS diffusion laws. It has to be
noted that variation of the domain size seen by ICS of FCS diffusion
laws shows strikingly similarities with heat capacity profile (for
comparison see Fig 8.6 p135 of \cite{Hei07})}}
      \label{fig8}
   \end{center}
\end{figure}

Using surface area of $a_{f} = 63 {\AA}^{2}$ and $a_{g} =
45{\AA}^{2}$ for PC lipids respectively in fluid phase or gel phase
\cite{Alm92,Wie89} allowed calculation of the mean radius of domains
in nm.  In Fig~\ref{fig8} are plotted the mean radius in nm of the
above defined domains obtained by FCS diffusion laws on the MC
simulation and compared to those viewed by ICS analysis or direct
space morphoanalysis. This comparison shows that the mean domains
radius obtained by our FCS diffusion laws are close to the one found
both by morphoanalysis and by ICS analysis. This clearly shows that
the $w^{2}_{0}$ is an indirect indication of the mean radius of the
domains.

\begin{figure}
   \begin{center}
      \includegraphics*[width=4in,height=3in]{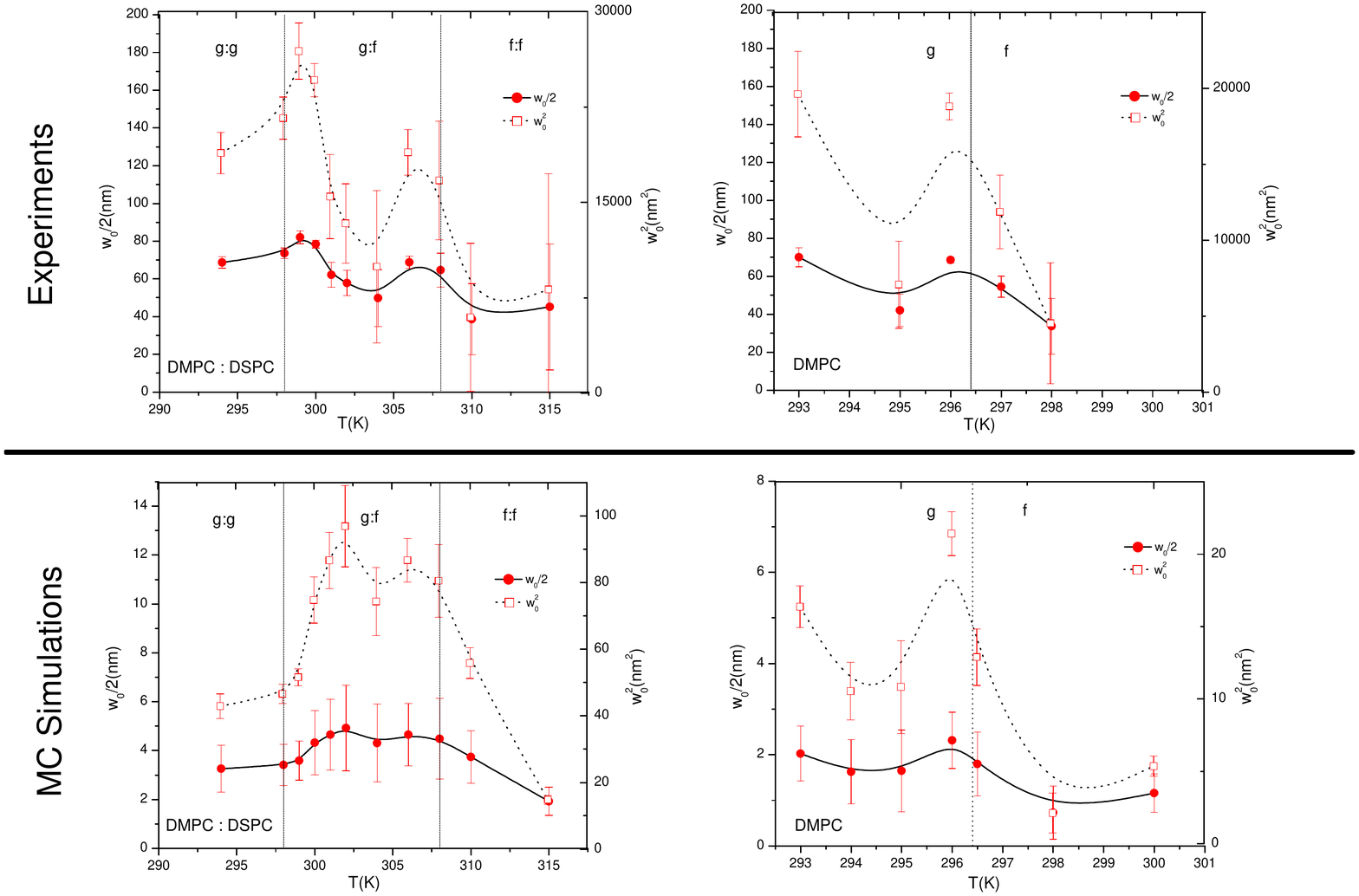}
      \caption{\small{\textbf{ Variation of $w^{2}_{0}$ and
$w_{0}/2$ as a function of temperature for MC simulated DMPC-DSPC
mixture or pure DMPC.}}\scriptsize{$w^{2}_{0}$ is estimated from the
linear fit using eq.~\ref{eqn:linfit} of the FCS diffusion law as
the value of $w^{2}$ for $\tau_{d}=0$. Here is depicted plots of
$w^{2}_{0}$ (right scale) and $w_{0}/2$ (left scale) as a function
of temperature determined experimentally on a DMPC:DSPC 8:2 mol:mol
mixture (Part A) or pure DMPC (Part B) or numerically for DMPC:DSPC
8:2 mol:mol (Part C) and pure DMPC (Part D). (\textit{Experimental
point are linked by a solid curve(for $\frac{w_{0}}{2}$) or dashed
curve (for $ w^{2}_{0}$)  polynomial fit to guide the eye.})}}
      \label{fig9}
   \end{center}
\end{figure}

Moreover, Fig~\ref{fig9}A, that shows changes in the experimental
$w^{2}_{0}$ and the mean radius ($w_{0}/2$) parameters found by FCS
diffusion laws as a function of temperature, exhibits a main change
between 301 and 302 K (close to the first melting temperature,
transition $gg:gf$) and a smaller shoulder around 306 K (close to
the second melting temperature, transition $gf:ff$). The shape of
the $w^{2}_{0}=f(T)$ curve is similar to the heat capacity profile
of the DMPC:DSPC (8:2 mol:mol). This is also seen in FCS diffusion
laws of DMPC alone (Fig~\ref{fig9}B). In that case, $w^{2}_{0}$
(equivalently $w_{0}/2$) show an increased at 296K, close to the
$g:f$ phase transition temperature of DMPC(\textit{ref}). This
variation of $w^{2}_{0}$ is also obtained in simulated FCS diffusion
laws of DMPC:DSPC mixture and DMPC alone respectively
(Fig~\ref{fig9}C and Fig~\ref{fig9}D). Whereas radius ($w_{0}/2$) of
domains viewed by FCS diffusion laws on MC simulations were found to
be less than 5 nm, experimental FCS diffusion laws shows values of
$w_{0}^{2}$ values between 10 000 and 26 000 nm$^{2}$ for the lipid
mixture, which finally give values of $w_{0}/2$ higher than 50 nm
and lower than 80 nm.

\section{Discussion}

Diffusion in this type of lipid mixture has been studied for a long
time by various techniques. Vaz \emph{et al.} performed FRAP
experiments \cite{Vaz89} on DMPC: DSPC 8:2 mol:mol at different
temperatures using NBD-DLPE as the fluorescent dye and found values
ranging from D=0.25 $\mu m^{2}.s^{-1}$ at T=293K to D=7.5 $\mu
m^{2}.s^{-1}$ at T=308K. The experimental FCS diffusion laws
obtained in this study exhibit value of diffusion coefficient
between 0.12 $\mu m^{2}.s^{-1}$ at 294K to 5.6 $\mu m^{2}.s^{-1}$ at
310K. Indeed, the global increase of D obtained by FCS diffusion
laws as a function of temperature is coherent with the one observed
using FRAP technique. In complex media, FRAP measurements generally
do not show total fluorescence recovery. This is described by
fitting FRAP recovery curves with a so-called immobile fraction.
This immobile fraction reflects the incapacity for one molecule to
leave the photodestructed area during the time of the experiment and
is therefore an indirect information on the restriction to the
diffusion. Immobile fraction is ranging between 0 (free diffusion)
and 1 (no diffusion). In their study, Vaz \emph{et al.} showed that
the immobile fraction varied from 0.52 at 293K to 0 at 308K with a
sharp transition in between, closely linked to the phases
transitions $gg:gl:ll$ of the lipid mixture \cite{Vaz89}. The FCS
diffusion laws observed in our study exhibit non zero and negative
$\tau_{d_{0}}$ below the second transition $gl:ll$. It has to be
pointed out that the shape of the evolution of the immobile fraction
as a function of the temperature of the lipid mixture is strikingly
similar to the evolution of our $\tau_{d_{0}}$ parameter.
%This is of
%interested since it is shown here that the $\tau_{d_{0}}$ parameter
%is linked to the fractional gel surface of the mixture and could
%therefore be used to estimate the relative amount of gel to liquid
%surface in the system as a function of temperature. To our
%knowledge, this approach has not been shown to be possible when
%analyzing the immobile fraction of FRAP experiments.
In a pure gel phase, one would expect to retrieve a free diffusion
behavior of the molecules. In this later case, the immobile
fraction, as well as our $\tau_{d_{0}}$ parameter should go back to
the zero value. This is not the case here, neither it is the case
for the immobile fraction in the study of Vaz \emph{et al.}
\cite{Vaz89}. One simple explanation could be that the system is not
in pure gel state at a temperature close to 294K. But the MC
simulation using the thermodynamics of the system shows that more
than 99$\%$ of the system is in the gel state. When this amount of
gel is present in the mixture, percolation is expected to occur and
to stop long range diffusion, which is the main diffusion component
seen in our FCS experiments. Nevertheless, it has been suggested
that even in gel state, ripple phase formation allows fast diffusion
along line defects \cite{Sch83}. This would lead to fast
fluctuations within the observed area and therefore a higher
diffusion coefficient than expected. For example, Hac \emph{et al.}
\cite{Hac05} also found using FCS a value of diffusion coefficient
in the gel phase in the range of 0.05 to 0.1 $\mu m^{2}.s^{-1}$
which is again significantly higher than the one reported by other
methods.
%Very long diffusion time are difficult to be correctly estimated by both techniques.
 Ripple phase formation is certainly not
the only explanation of these high diffusion coefficients found in
the gel phase since our MC simulations exhibit exactly the same
behavior regarding the $\tau_{d}^{0}$ parameter as experiments do,
regardless the fact that ripple phases do not exist in the
simulations. Lipid state change could also be a source of
fluorescence fluctuations within a given area and therefore lead to
a new correlation time. The system studied here exhibits three
different time scales : one characteristic of diffusion in a pure
liquid environment ($\tau^{l}_{d}$), one in a pure gel environment
($\tau^{g}_{d}$) and finally one proportional to the rate of state
change from liquid to solid (or inversely) for each lipid chains. It
has been shown that this latter time has an influence on the
apparent diffusion \cite{Hac05,Sug05}. If state changes occur more
rapidly than the time needed for a tracer to diffuse into a gel
obstacle, the obstacle itself will be able to fluctuate and to
change his shape and size. This is equivalent to diffusion in smooth
obstacle where it has been shown that percolation has a much smaller
effect than expected \cite{Tor91}. Finally, time scale of the method
is also very important. FCS or FRAP experiments are conducted over
tenths of seconds which does not authorize, at diffraction limited
spots, to measure D lower than $10^{-4}$ or $10^{-5} \mu
m^{2}.s^{-1}$. All these possibilities can explain the finding of a
higher diffusion coefficient than the one intuitively expected and a
non zero $\tau_{d}^{0}$ in the pure gel phase at 294K.

Another parameter of interest in our FCS diffusion laws is the value
of the intercept of asymptotic linear extrapolation of the diffusion
law with the abscissa axis. This parameter is empirically defined
has the "zero diffusion time" characteristic waist (value of $w^{2}$
at $\tau_{d}=0$, $w^{2}_{0}= \frac{-\tau_{d_{0}}}{4D_{eff}}$). This
empirical parameter has been analysed has a function of temperature.
It exhibits a very good correlation to the differential scanning
calorimetry curve with two maxima located at the same temperature
(1K accuracy) for DMPC:DSPC 8:2 mol:mol and one maximum at 296K for
DMPC alone. Liquid or solid domains are known to be maximum in size
at the maximal enthalpy \cite{Hei98, Ebe01}, it seemed therefore
reasonable to analyse this $w^{2}_{0}$ parameter as being a function
of the domain size. Wawrezinieck \textit{et al} \cite{Waw05} have
shown by using classical random walk simulations, that in given
geometries, this parameter allows to estimate the exact size of the
mesh or the domain. These numerical simulations results have been
confirmed by analytical solving of the problem by N. Destainville
\cite{Des08}. The geometry of the system studied here is totally
unknown and depends only on the thermodynamic properties of it.
Therefore it seems very difficult to obtain an analytical model that
precisely define the size of the domains as a function of
$w^{2}_{0}$. Nevertheless, by using ICS or morphometry analysis of
the images obtained from the MC simulation, it is shown here that
the radius of circular-like domains found in the images is close to
the size of those estimated by means of the $w^{2}_{0}$ parameter
obtained from the FCS diffusion laws. This clearly shows that
$w^{2}_{0}$ is a good estimator of the size of the domains in the
lipid mixture. In this study, circular-like domains are found to
measure between 2.5 to 5 nm in radius in the case of MC simulations
while they exhibit size in between 50 to 80 nm in experiments. AFM
studies on a 50:50 mol:mol DMPC:DSPC supported bilayers lipid
mixture has shown to exhibit domain size between 50 (circular) and
150x300 nm (rectangular) in the $g:f$ phase coexistence
\cite{Gio01}. These values are coherent with the one observed here
experimentally for our 80:20 mol:mol DMPC:DSPC lipid mixtures. Gliss
\emph{et al.} \cite{Gli98} also showed existence of nanoscale
domains in DMPC-DSPC mixture. AFM imaging allowed them to measure
gel domains of 10 nm, while neutron diffraction permitted
identification of gel domains exhibiting values of 7 nm. These size
are comparable to the one observed here in our MC simulations
whether by ICS analysis or by estimation of the $w^{2}_{0}$
parameter in the simulated FCS diffusion laws.

As stated above, this two-phases two-components DMPC:DSPC lipid
mixture has been studied for a long time by many different
techniques both on dynamic and structural approaches. Bagatolli
\emph{et al.} \cite{Bag00} even showed existence of micrometer size
domains within GUVs made of DMPC:DSPC 1:1 mol:mol.

Indeed, all these observations show that such systems exhibit
different space and time scale confinement, rising the question of
the anomality of the diffusion. Anomalous diffusion has been
extensively used to describe and study many different type of
dynamic behavior of biological molecules in their complex media (for
review see \cite{Dix08}). Basically, it can be seen as a general
case for diffusion. Relation between time and space could be
generalized as $<r^{2}>\propto t^{\alpha}$, with $0<\alpha<2$. If
$\alpha=1$, the diffusion is normal and described by pure Brownian
motion whereas if $\alpha<1$ the system is sub-diffusive indicating
that the molecular motions are restricted. DMPC:DSPC two-phases
two-components lipid mixture has been described by means of
anomalous diffusion both on MC simulation \cite{Sug05} and
experimentally \cite{Hac05}. In this case, $\alpha$ was found to be
systematically less than one, except in the $f:f$ region and in the
far $g:g$ region (very low temperature), where the molecular motion
becomes Brownian again. FCS diffusion laws obtained on both MC
simulations and experiments have also been fitted using anomalous
diffusion. In this study, $\alpha$ has been found to vary between
0.64 and 1 in both cases (MC simulations and experiments) exhibiting
two minima at the phase transitions $g:g\rightarrow g:f$ and
$g:f\rightarrow f:f$  (data not shown). Anomalous diffusion can
therefore clearly describe the molecular motion in this two-phases
two-components lipid mixtures or during phase transition in a pure
lipid, nevertheless, even if it gives informations on the
heterogeneity of the systems, it fails to quantitatively interpreted
the structure of this heterogeneous medium. It has been a deliberate
choice all along this work to compare results at extremely low
spatial scale (MC simulation) with experimental scale, this later
being 3 orders of magnitude larger. This is motivated by anomalous
properties of diffusion in the lipid mixture that exhibit similar
space and time behavior at any scale. This question of time to space
invariance can also be viewed \emph{a posteriori} by the consistency
of the results found here ($w_{0}$ versus T for example) although
simulations and experiments are compared at very different scales.

\section{Conclusion}

Here is presented a study of the two-phases two-components DMPC:DSPC
8:2 mol:mol lipid mixture and pure DMPC lipid system by means of FCS
diffusion laws both experimentally and using MC simulation with a
complete thermodynamic description. It is clearly shown that these
FCS diffusion laws, allows quantitative characterization of the
system in terms of diffusion, phase transition and mean size of the
gel or fluid domains presents in the lipid mixture. Domain size and
transient confinement times has already been shown to be predictable
by the use of FCS diffusion laws on defined geometries and pure
random walk model \cite{Waw05}. Here it is shown that this can be
generalized to a more complex model where the geometry is unknown
and the molecular motion are only driven by the thermodynamic
parameters of the system itself. Finally, the parameter obtained
from the interpretation of the FCS diffusion laws issued of the
numerical simulation are shown to have the same behavior than the
one obtained on experimental system. This confirms the advantages of
using FCS diffusion laws on experimental systems to described their
temporal and spatial structure.

% ----------------------------------------------------------------
\bibliographystyle{unsrt}
\bibliography{FCSlipids}

\clearpage
\begin{figure}
   \begin{center}
      \includegraphics*[width=5in]{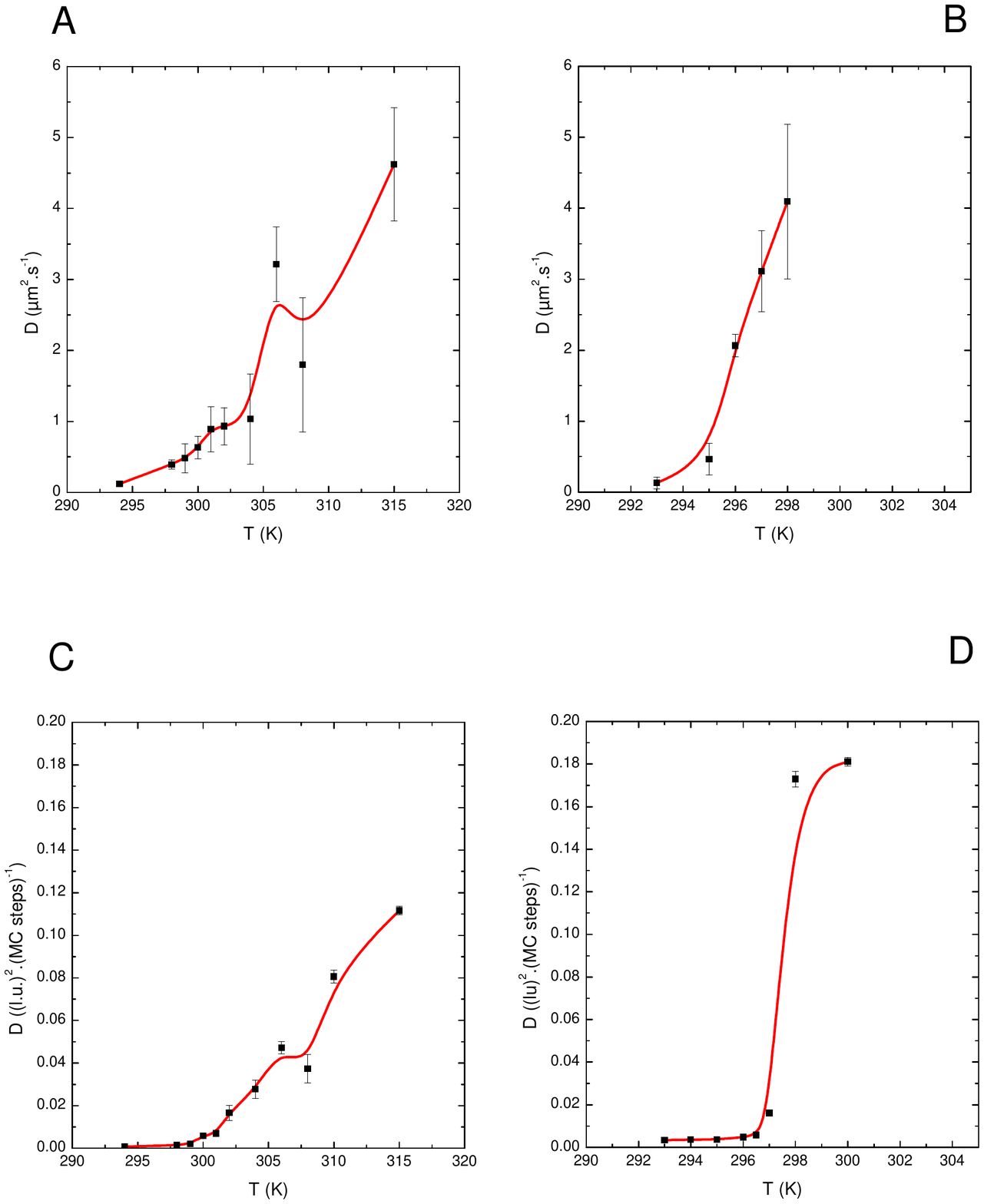}
      \caption{\small{\textbf{Effective diffusion coefficient
obtained by FCS diffusion laws.}}\scriptsize{Variation of the
diffusion coefficient ($D_{eff}$) obtained by measuring the slope of
the different fits using eq.~\ref{eqn:linfit} of the FCS diffusion
laws as a function of temperature. This clearly show, as expected,
an increase in $D$ with temperature from a lower plateau (gel phase)
to a higher one (liquid phase). Part A and B are values
experimentally obtained whereas part C and D are numerically
obtained for DMPC:DSPC 8:2 mol:mol and pure DMPC respectively.}}
      \label{fig1s}
   \end{center}
\end{figure}

\end{document}